\documentstyle[epsf,preprint,eqsecnum,aps]{revtex}
\begin{document}
\draft
\title{Theory of Cylindrical Tubules and Helical Ribbons
of~Chiral~Lipid~Membranes}
\author{J. V. Selinger,$^1$ F. C. MacKintosh,$^2$ and J. M. Schnur$^1$}
\address{$^1$Center for Bio/Molecular Science and Engineering,
Naval Research Laboratory, Code 6900, \\
4555 Overlook Avenue, SW, Washington, DC 20375 \\
$^2$Department of Physics, University of Michigan, Ann Arbor, MI 48109-1120}
\date{November 22, 1995}
\maketitle

\begin{abstract}
We present a general theory for the equilibrium structure of cylindrical
tubules and helical ribbons of chiral lipid membranes. This theory is based
on a continuum elastic free energy that permits variations in the direction
of molecular tilt and in the curvature of the membrane. The theory shows
that the formation of tubules and helical ribbons is driven by the chirality
of the membrane. Tubules have a first-order transition from a uniform state
to a helically modulated state, with periodic stripes in the tilt direction
and ripples in the curvature. Helical ribbons can be stable structures, or
they can be unstable intermediate states in the formation of tubules.
\end{abstract}


\narrowtext

\section{Introduction}

\begin{figure}
\centering\leavevmode\epsfysize=7.9in\epsfbox{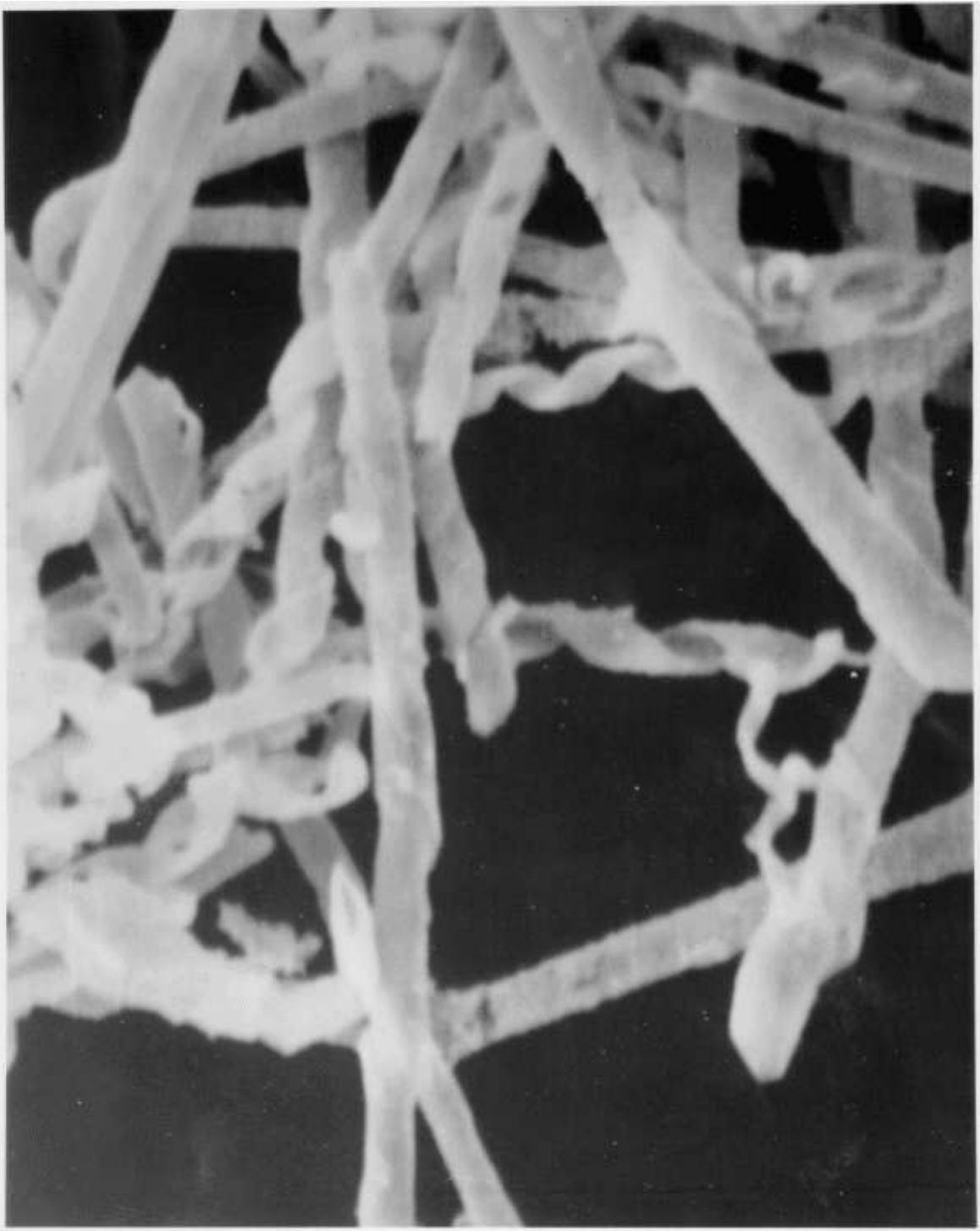}
\caption{Scanning electron micrograph of helical ribbons formed from a
racemic mixture of D and L diacetylenic phospholipids.  Note that the
racemic mixture forms distinct right- and left-handed helical ribbons.
The diameter of the helical ribbons is approximately 0.5~$\mu$m.
Reprinted from Ref.~\protect\cite{singh} with permission \copyright
1988 Elsevier Scientific Publishers.}
\label{helixphoto}
\end{figure}

Chiral amphiphilic molecules can self-assemble into microstructures with
a variety of morphologies. Two of the most interesting morphologies, both
for basic research and for technological applications, are cylindrical
tubules and helical ribbons~\cite{yagerschoen,joelscience,acsbook}.
Tubules are bilayer or multilayer membranes of amphiphilic molecules
wrapped in a cylinder, as shown in Fig.~\ref{tubulephoto}~\cite{gulik}.
They are observed in a variety of systems, including diacetylenic
lipids~\cite{yagerschoen,markowitz1}, bile~\cite{chung},
surfactants~\cite{tachibana}, and glutamates~\cite{nakashima}. In
diacetylenic lipid systems, the tubule diameter is typically 0.5~$\mu $m
and the tubule length is typically 50--200~$\mu $m; in other systems, the
tubule dimensions are typically several times larger~\cite{chung}.
In most of these systems,
tubules exhibit a helical ``barber-pole'' pattern on the surface of the
cylinder. Helical ribbons are similar microstructures, consisting of long
twisted strips of membranes with their edges exposed to the solvent, as
shown in Fig.~\ref{helixphoto}~\cite{singh}. In some cases, helical
ribbons are unstable precursors to the formation of tubules; in other
cases, helical ribbons appear to be stable. Cylindrical tubules have been
studied extensively for use in several technological applications, such
as electroactive composites and controlled-release
systems~\cite{joelscience}. Helical ribbons have not been used in
technological applications, but they have also been studied extensively
as part of an effort to rationally control the self-assembly of tubules.

There have been three general approaches to the theory of tubules and
helical ribbons. First, de Gennes argued that a membrane of chiral
molecules in any tilted phase must develop a spontaneous electrostatic
polarization~\cite{degennes}. This polarization can induce a narrow strip
of membrane to buckle into a cylinder. The original theory of de Gennes
described buckling along the long axis of a cylinder, but a
straightforward modification describes helical winding due to
electrostatic interactions. This theoretical approach predicts that
adding of electrolytes to the solvent should increase the radius of the
resulting tubules, because the electrostatic interaction would be
screened by electrolytes in solution. However, experiments have shown
that electrolytes in solution do not affect the formation or radius of
tubules~\cite{chappell1}, except for the particular case of tubules
composed of amphiphiles with charged head groups~\cite{markowitz2}. Thus,
electrostatic interaction is very probably
not a dominant factor in tubule formation.

As an alternative theoretical approach,
Lubensky and Prost derived a general
phase diagram for membranes with in-plane orientational order, which
predicts cylinders as well as spheres, flat disks, and
tori~\cite{lubenskyprost}. Within the cylindrical phase, the cylinder
radius and length are determined by a competition between the curvature
energy and the edge energy. One important prediction of this theory is
the scaling $r\propto\ell^{1/2}$ between the tubule radius $r$ and
the tubule length $\ell$. However, experiments have found no
correlation between $r$ and $\ell$. Rather, in a typical sample of
tubules, $r$ is quite monodisperse while $\ell$ varies
widely~\cite{georger}. Thus, the competition between curvature energy and
edge energy is also not a dominant factor in tubule formation.

A third approach, which is consistent with the experimental results
on lipid tubules, is
based on the chiral packing of the molecules in a membrane. Helfrich and
Prost have shown that a chiral membrane in a tilted phase will form a
cylinder because of an intrinsic bending force due to
chirality~\cite{helfrichprost}. This bending force arises because long
chiral molecules do not pack parallel to their neighbors, but rather at a
nonzero twist angle with respect to their neighbors. If the molecules lie
in bilayers, and are tilted with respect to the local layer normal, the
favored twist from neighbor to neighbor leads the whole membrane to twist
into a cylinder.

Several investigators have generalized the original Helfrich-Prost
concept of an intrinsic chiral bending force in different ways. Ou-Yang
and Liu have developed a version of this theory based on an analogy with
cholesteric liquid crystals~\cite{ouyang}. Chappell and Yager have
developed an analogous theory in which the direction of one-dimensional
chains of molecules, rather than the direction of molecular tilt, defines
a vector order parameter within the membrane~\cite{chappell2}. Nelson and
Powers have used the renormalization group to calculate the effects of
thermal fluctuations on tubules~\cite{nelson}. This calculation predicts
an anomalous scaling of the tubule radius as a function of the strength
of the chiral interaction. Chung {\it et al.} have considered the full
elastic anisotropy of a membrane, and have related the pitch angle of
tubules to a ratio of membrane elastic constants~\cite{chung}. They have
also made predictions for the kinetic evolution of helical ribbons into
tubules.

\begin{figure}
\epsfbox{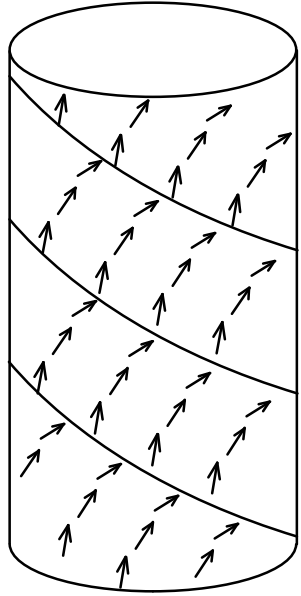}
\caption{Schematic view of the striped pattern in the tilt direction in
the modulated state of a tubule.  The arrows indicate the direction of
the molecular tilt, projected into the local tangent plane.}
\label{tiltmodulation}
\end{figure}

In our earlier paper, we generalized the Helfrich-Prost concept in
another way~\cite{ss}.  In that theory, we began with a three-dimensional
(3D) liquid-crystal free energy and applied it to the case of chiral
molecules in a membrane.  We showed that the chiral term in this free
energy has two simultaneous effects:  it leads the membrane to twist into
a cylinder, and it induces a variation in the direction of the molecular
tilt on the cylinder.  Through this calculation, we predicted that
tubules can form periodic modulated structures characterized by helical
stripes in the tilt direction winding around the cylinders, as shown in
Fig.~\ref{tiltmodulation}.  These stripes are analogous to the stripes
seen in thin films of chiral smectic liquid
crystals~\cite{LangerSethna,Hinshaw,jacobs}, but in a cylindrical rather
than a planar geometry.  We argued that these stripes correspond to the
helical substructure that is often observed on tubules.

\begin{figure}
\epsfbox{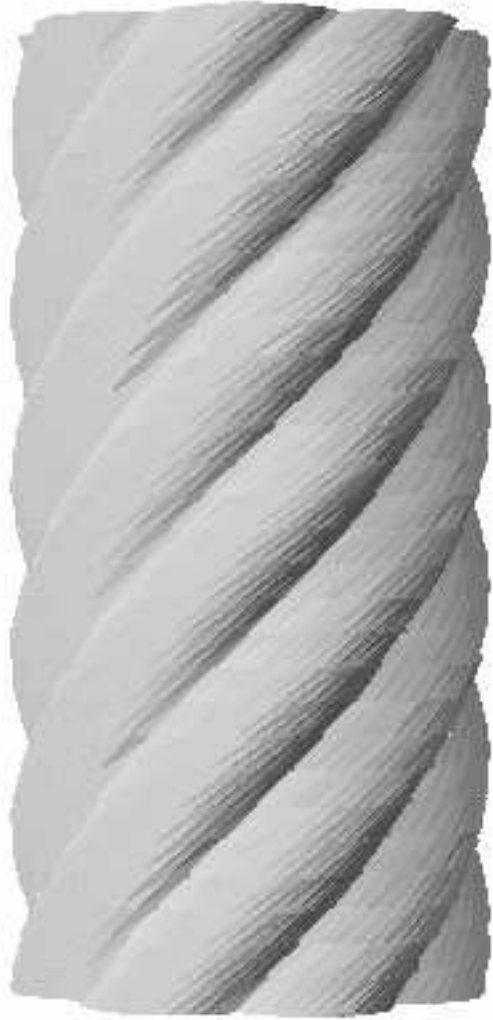}
\caption{Schematic view of the ripples in the curvature in the modulated
state of a tubule.  The amplitude of the ripples has been exaggerated for
clarity.}
\label{ripple3D}
\end{figure}

In this paper, we further extend this theoretical approach to provide a
more unified and systematic model of tubules and helical ribbons. First,
we combine our earlier theory with the Chung theory to obtain a theory of
tubules with both elastic anisotropy and tilt modulation. We derive
explicit expressions for the average direction of the tilt and for the
direction of the modulation, in terms of the elastic coefficients, and we
show that these directions are different. Next, we show that the same
theory can also be applied to helical ribbons. This theory shows that
helical ribbons can be stable microstructures for a certain range of
parameters; they are not necessarily intermediate states in the formation
of tubules. Finally, we investigate the effects of the striped tilt
modulation on the detailed shape of tubules. Based on recent models of
lipid bilayers~\cite{Pbeta1,Pbeta2}, we expect a modulation in the tilt
to induce a modulation in the curvature. We predict the curvature
modulation illustrated in Fig.~\ref{ripple3D}, which resembles the
$P_{\beta ^{\prime }}$ rippled phases of lyotropic liquid
crystals~\cite{Luzzati,Joe}.

The plan of this paper is as follows. In Sec.~II, we construct a free
energy for tubules and helical ribbons, which includes elastic anisotropy
and includes the possibility of tilt modulation. In Sec.~III, we use this
free energy to predict the radius and tilt direction for tubules with
uniform tilt. The results of that section are in exact agreement with the
theory of Chung {\it et al.} In Sec.~IV, we consider tubules with a
modulation in the tilt. We propose an explicit ansatz for that
modulation, and then minimize the free energy over the parameters in that
ansatz. We find that there is a first-order transition from uniform to
modulated tubules. In Sec.~V, we apply the same theory to helical
ribbons. The theory predicts the range of parameters in which helical
ribbons are stable. In Sec.~VI, we investigate ripples in the curvature
of tubules and helical ribbons. We derive an explicit expression for the
ripple shape that corresponds to our ansatz for the tilt modulation.
Finally, in Sec.~VII, we discuss these theoretical predictions and
compare them with experimental results.

\section{Free Energy}

In this section, we construct a free energy for tubules and helical
ribbons. In this free energy, we suppose that the membrane is in a fluid
phase, which may have hexatic bond-orientational order but does not have
crystalline positional order. This assumption is based on the theoretical
argument of Nelson and Peliti, who showed that a flexible membrane cannot
have crystalline positional order in thermal equilibrium at nonzero
temperature, because thermal fluctuations induce dislocations, which
destroy this order on long length scales~\cite{nelsonpeliti}. This
assumption is also supported by two types of experimental evidence.
First, Treanor and Pace found a distinct fluid character in nuclear
magnetic resonance and electron spin resonance experiments on
tubules~\cite{treanor}. Second, Brandow {\it et al.} found that tubule
membranes can flow to seal up cuts in the membrane from an atomic force
microscope tip~\cite{brandow}. This flow indicates that the membrane has
no shear modulus~\cite{sackmann}. Further information on the membrane
phase comes from Thomas {\it et al.,} who found long positional
correlation lengths of 0.068 to 0.135~$\mu$m in synchrotron x-ray
scattering studies of tubules~\cite{brittscience}. Taken together, all
of these experimental results suggest that tubule membranes are in a
highly correlated hexatic phase.

\begin{figure}
\epsfbox{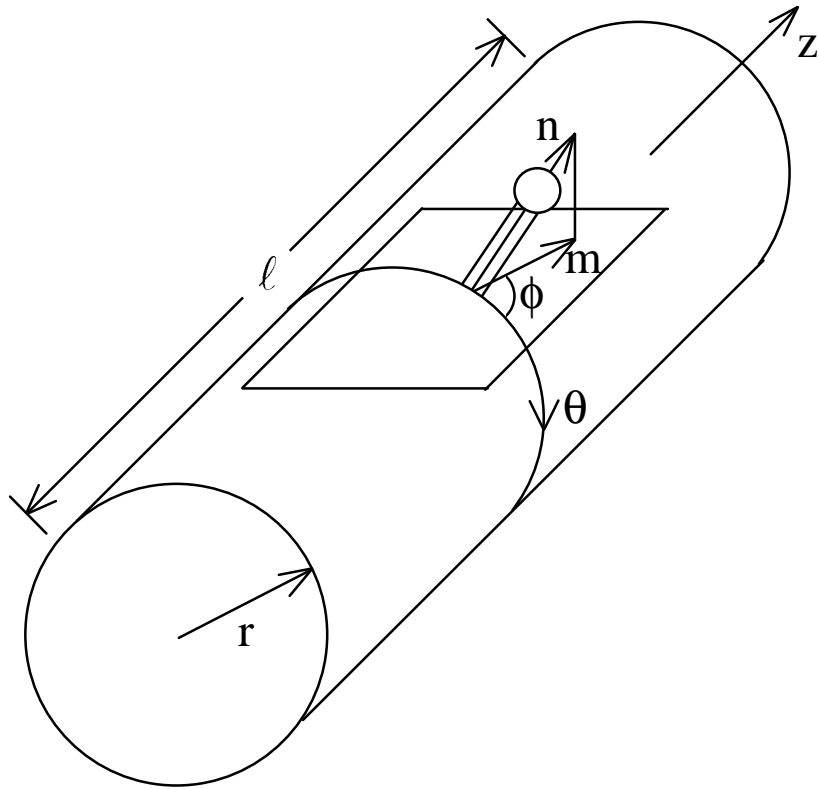}
\caption{Geometry of a tubule with radius $r$ and length $\ell$.  Here,
$\vec n$ is the molecular director, $\vec m$ is the projection
of $\vec n$ into the local tangent plane (normalized to unit magnitude),
and $\phi$ is the angle in the tangent plane between $\vec m$ and the
equator.}
\label{phidef}
\end{figure}

In this paper, we do not attempt to derive the free energy from a 3D
liquid-crystal free energy, as in our earlier paper~\cite{ss}. Rather, we
use a 2D differential-geometry notation, which is more general.  In this
notation, the curvature of a membrane is described by the curvature
tensor $K_{ab}$.  The orientation of the molecular tilt is represented by
the unit vector $\vec m$. As shown in Fig.~\ref{phidef}, $\vec m$ is the
projection of the molecular director $\vec n$ into the local tangent
plane, normalized to unit magnitude.

A membrane may consist of a single domain, in which the curvature
$K_{ab}$ is low and the tilt direction $\vec m$ varies smoothly.
Alternatively, a membrane may consist of many internal domains
separated by domain walls.  Domain walls are sharp boundaries between
domains.  Across a narrow domain wall, $\vec m$ jumps abruptly and
$K_{ab}$ may become high.  Here, we will consider the free energy of
domains and domain walls separately.

First, we consider a domain of low curvature and smoothly varying tilt.
Within a domain of a membrane, the free energy can be written as
\begin{eqnarray}
F=\int dA  \biggl[&&\frac{1}{2}\kappa\left(K^a_a\right)^2
+\frac{1}{2}\kappa' m^a m^b K_{ac} K^c_b\nonumber\\
&&+\lambda_{\rm HP} m^a m^c K^b_a \sqrt{g}\epsilon_{bc}\nonumber\\
&&-\lambda_{\rm LS} \sqrt{g}\epsilon_{ab} D^a m^b
-\gamma K^a_a D_b m^b\nonumber\\
&&+\frac{1}{2} c \left[\left(\sqrt{g}\epsilon_{ab} D^a m^b\right)^2
+\left(D_a m^a\right)^2\right]\biggr].
\label{freegeneral}
\end{eqnarray}
Here, the surface area element is $dA=\sqrt{g}d^2x$, $\epsilon _{ab}$ is
the antisymmetric symbol, and $D_a$ is the covariant derivative within
the membrane. Throughout this work, we shall imply summation over
repeated indices, such as $a$ in the first term above. The first term of
$F$ is the standard Helfrich bending energy of the
membrane~\cite{helfrich}. The coefficient $\kappa $ is the isotropic
rigidity. The second term represents the anisotropy of the rigidity. The
coefficient $\kappa ^{\prime }$ is the difference between the energy
required to bend the membrane parallel to the tilt $\vec m$ and the
energy required to bend it perpendicular to $\vec m$. In general,
$\kappa^{\prime }$ can be either positive or negative. The third term,
introduced by Helfrich and Prost~\cite{helfrichprost}, is a chiral term
that favors curvature in a direction $45^{\circ }$ from $\vec m$. The
coefficient $\lambda _{{\rm HP}}$ is a measure of the magnitude of the
chiral interaction between the molecules in the membrane. The fourth term
is another chiral term, which favors a bend in $\vec m$. This term was
introduced by Langer and Sethna in the context of flat films of chiral
liquid crystals~\cite{LangerSethna}. The coefficient
$\lambda _{{\rm LS}}$ should have the same magnitude as
$\lambda _{{\rm HP}}$; both coefficients are comparable to $Kq$ in the
notation of our earlier paper~\cite{ss}. However, there is no symmetry
reason why $\lambda _{{\rm LS}}$ and $\lambda _{{\rm HP}}$ must be equal.
The fifth term, with coefficient $\gamma $, is the coupling between
curvature of the membrane and splay of $\vec m$. It can be understood in
two equivalent ways: (a)~the curvature of the membrane breaks the
symmetry between the two monolayers of a bilayer and hence induces a
splay in $\vec m$, or (b)~a splay in $\vec m$ induces a curvature of the
membrane to keep the molecules more nearly parallel in three dimensions.
This term is equivalent to terms that have been considered in the
Lubensky-MacKintosh theory of $P_{\beta ^{\prime }}$ rippled
phases~\cite{Pbeta1,Pbeta2}. The final pair of terms is the 2D Frank free
energy for variations in $\vec m$. It provides an energy penalty for tilt
variations, and hence limits those variations. The coefficient $c$ is a
single Frank constant.

We now consider a domain wall between two domains.  Across a domain wall,
the tilt direction $\vec m$ changes abruptly.  The curvature $K_{ab}$ may
also become large, giving a ``crease'' in the membrane.  For those
reasons, different parts of the molecules come into contact at a domain
walls than inside the domains.  A domain wall therefore costs an energy
$\epsilon _W$ per unit length.  The domain-wall energy cannot be derived
from the free energy~(\ref{freegeneral}) within a domain; rather, it is
an additional parameter that describes narrow walls where that free
energy does not apply. For any morphology that includes domain walls,
with a total wall length $L_W$, the total free energy is
$F_{{\rm tot}}=F+\epsilon _W L_W$.

In the free energy~(\ref{freegeneral}), the term
$\lambda _{{\rm LS}}\sqrt{g}\epsilon _{ab}D^a m^b$ is a total derivative.
If the curvature is constant, as in a perfect cylinder, the term
$\gamma K_a^a D_b m^b$ is also a total derivative. One might think that
these terms integrate to constants and hence do not affect the morphology
of a membrane. However, these terms are significant if the membrane
consists of  internal domains separated by domain walls, because they
then become line integrals along the domain walls. Indeed, these terms
favor the formation of a series of domains separated by domain walls,
because they can contribute a negative free energy for each domain. Thus,
in the rest of this paper, we will retain these terms, and we will
explicitly compare their negative contribution to the free energy with
the energy cost of the domain walls.

The free energy~(\ref{freegeneral}) does not contain an exhaustive list
of all the terms permitted by symmetry; other terms are also possible.
For example, the term $(m^am^bK_{ab})^2$ would give an additional elastic
anisotropy, which has been considered by Chung {\it et al.}~\cite{chung}.
The combination $(\sqrt{g}\epsilon _{ab}D^am^b)^2-(D_bm^b)^2$ would give
a difference in the Frank constants for bend and splay. Rather, this free
energy is only intended to include the terms that generate distinct
physical effects, while the omitted terms give higher-order details of
the structure.

If the membrane forms a perfect cylinder with radius $r$, the free energy~(%
\ref{freegeneral}) simplifies greatly. In the standard cylindrical
coordinates $(\theta ,z)$, the curvature tensor becomes
\begin{equation}
K=\left(
\begin{array}{cc}
-1/r & 0 \\
0 & 0
\end{array}
\right) .
\end{equation}
The tilt director field can be written as
$\vec m(\theta,z)=(\cos\phi,\sin\phi)$. As shown in Fig.~\ref{phidef},
$\phi(\theta,z)$ is the angle in the local tangent plane between the tilt
direction and the equator of the cylinder. The free energy then becomes
\begin{eqnarray}
F=\int dA \biggl[&&\frac{1}{2}\kappa\left(\frac{1}{r}\right)^2
+\frac{1}{2}\kappa' \left(\frac{1}{r}\right)^2 \cos^2 \phi\nonumber\\
&&-\lambda_{\rm HP} \left(\frac{1}{r}\right) \sin\phi\cos\phi\nonumber\\
&&-\lambda_{\rm LS} \vec\nabla\times \vec m
+\gamma \left(\frac{1}{r}\right) \vec\nabla\cdot \vec m\nonumber\\
&&+\frac{1}{2} c \left[\left(\vec\nabla\times \vec m\right)^2
+\left(\vec\nabla\cdot \vec m\right)^2\right]\biggr],
\label{free1}
\end{eqnarray}
where $dA=rd\theta dz$,
\begin{equation}
\label{delcross}\vec \nabla \times \vec m=\frac 1r\partial _\theta \sin \phi
-\partial _z\cos \phi ,
\end{equation}
and
\begin{equation}
\label{deldot}\vec \nabla \cdot \vec m=\frac 1r\partial _\theta \cos \phi
+\partial _z\sin \phi .
\end{equation}
Note that the 2D cross product defined by Eq.~(\ref{delcross}) is a scalar.
This form of the free energy is generally more convenient to work with than
the more general form~(\ref{freegeneral}). The interpretation of these terms
is exactly as discussed above. In the following sections, we will use
this free energy to investigate the structure of tubules and helical
ribbons.

\section{Tubules with Uniform Tilt}

As a first step in understanding the implications of the free
energy~(\ref{freegeneral}), we consider tubules with a uniform tilt
direction $\vec m$. This model of uniform tubules is equivalent to the
model of Chung {\it et al.}~\cite{chung}. We present it here in order to
review their results in our notation.

\begin{figure}
\epsfbox{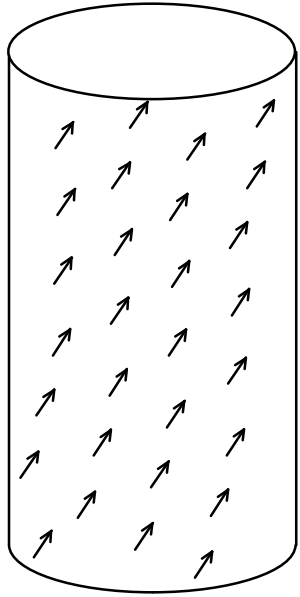}
\caption{Schematic view of a tubule with the uniform tilt direction
$\vec m=(\cos\phi_0,\sin\phi_0)$, as indicated by the arrows.}
\label{uniformtilt}
\end{figure}

Suppose that a tubule of radius $r$ has a uniform tilt
$\vec m=(\cos\phi_0,\sin\phi_0)$, at an angle $\phi_0$ away from the
equator, as shown in Fig.~\ref{uniformtilt}. The entire tubule is a
single domain, with no domain walls. In terms of $r$ and $\phi_0$,
the free energy per unit area is
\begin{eqnarray}
\frac{F}{A}=&&\frac{1}{2}\kappa\left(\frac{1}{r}\right)^2
+\frac{1}{2}\kappa' \left(\frac{1}{r}\right)^2 \cos^2 \phi_0\nonumber\\
&&-\lambda_{\rm HP} \left(\frac{1}{r}\right) \sin\phi_0\cos\phi_0.
\end{eqnarray}
Minimizing this expression over $r$ and $\phi_0$ simultaneously, we obtain
\begin{equation}
\label{rmin}r=\frac{\kappa^{1/4}\left(\kappa+\kappa^{\prime}\right)^{1/4}
\left[\kappa^{1/2}+ \left(\kappa+\kappa^{\prime}\right)^{1/2}\right]}%
{\lambda_{{\rm HP}}}
\end{equation}
and
\begin{equation}
\label{phi0min}\phi_0=\arctan\left[\left(\frac{\kappa+\kappa^{\prime}}%
{\kappa}\right)^{1/4}\right].
\end{equation}

These results lead to two important conclusions. First, the tubule radius $r$
scales inversely with the chirality parameter $\lambda_{{\rm HP}}$. In a
nonchiral membrane, where $\lambda_{{\rm HP}}\rightarrow0$, the radius $%
r\rightarrow\infty$. The coefficient of $1/\lambda_{{\rm HP}}$ is a
combination of the elastic constants $\kappa$ and $\kappa^{\prime}$. Second,
the tilt direction $\phi_0$ is determined by the ratio of the energy cost $%
\kappa+\kappa^{\prime}$ for bend parallel to the tilt direction over the
energy cost $\kappa$ for bend perpendicular to the tilt direction. In an
isotropic membrane, where $\kappa^{\prime}=0$, we obtain $\phi_0=45^\circ$.
In an anisotropic membrane, $\kappa^{\prime}$ may be positive or negative,
and hence $\phi_0$ may be greater or less than $45^\circ$.

Chung {\it et al.}~\cite{chung} originally derived Eq.~(\ref{phi0min}) in
order to analyze experimental data on tubules in bile. In experiments on
bile systems, they found two types of tubules. Both types of tubules show
clear helical markings, but the direction of these markings are
different: $53.7^\circ$ from the equator in one type of tubule and
$11.1^\circ$ from the equator in the other. Chung {\it et al.} assumed
that these helical markings are aligned with the molecular tilt, so that
$\phi_0=53.7^\circ$ and $11.1^\circ$ in the two types of tubules. They
then used Eq.~(\ref {phi0min}) to determine the values of
$(\kappa+\kappa^{\prime})/\kappa$ that correspond to these values of
$\phi_0$. The resulting values are $(\kappa+\kappa^{\prime})/\kappa=3.4$
and 0.0015, respectively. The ratio of 3.4 is quite reasonable, but the
ratio of 0.0015 is surprising. It seems unlikely that a bend in the
membrane perpendicular to the tilt direction would cost almost 1000 times
more energy than a bend parallel to the tilt direction. For that reason,
we must re-examine the assumption that helical markings are aligned with
the molecular tilt. In the following section, we propose an alternative
interpretation of the helical markings, which can explain this anomaly.

\section{Tubules with Tilt Modulation}

In this section, we relax the assumption that the tilt direction $\vec m$ is
uniform everywhere on a tubule. Instead, we allow $\vec m$ to vary as a
function of the cylindrical coordinates $\theta$ and $z$, with $\vec m%
(\theta,z)=(\cos\phi(\theta,z),\sin\phi(\theta,z))$. We show that the ground
state of the tubule can have a periodic helical modulation of $\vec m$.

\begin{figure}
\epsfbox{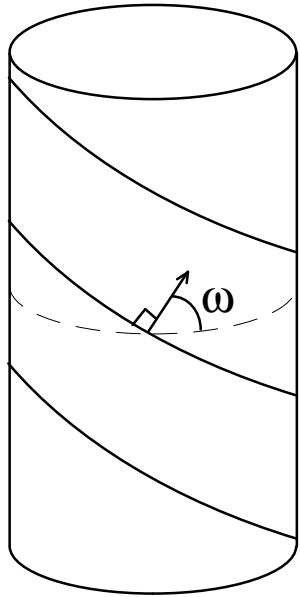}
\caption{Definition of the angle $\omega$, which gives the direction of
the stripes and the domain walls on a tubule with respect to the equator.}
\label{omegadef}
\end{figure}

\begin{figure}
\epsfbox{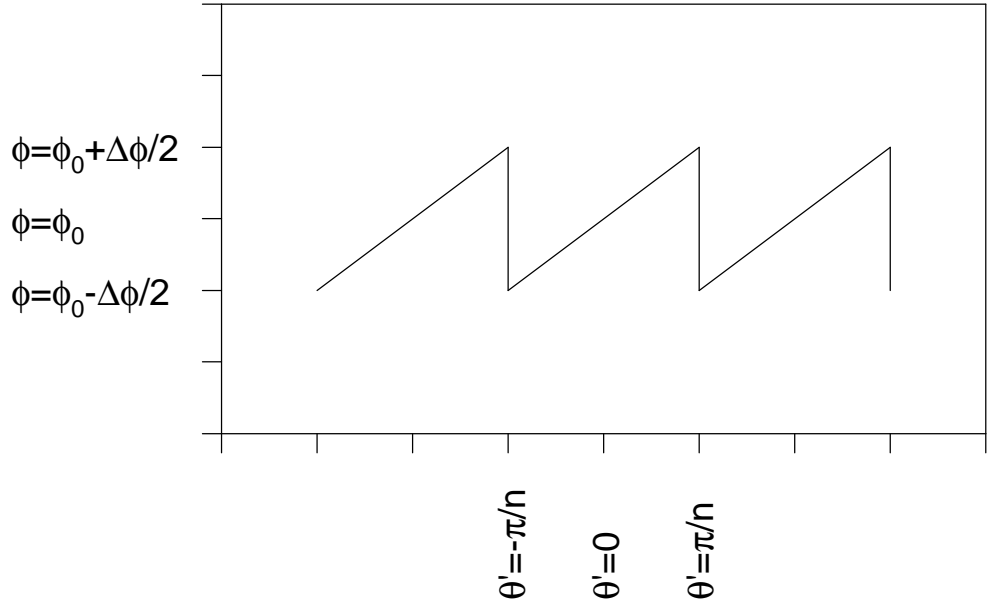}
\caption{Ansatz for the tilt direction $\phi(\theta,z)$, in terms of the
coordinate $\theta'\equiv\theta+(z/r)\tan\omega$.}
\label{sawtooth}
\end{figure}

As a specific ansatz, we suppose that there are $n$ distinct stripes in
the tilt direction.  In the terminology of Sec.~II, each stripe is a
single domain of the membrane.  The stripes are separated by $n$ distinct
domain walls. The stripes and domain walls run around the cylinder
helically, as shown in Fig.~\ref{tiltmodulation}. We further suppose that
the tilt direction $\phi $ varies linearly across each stripe. Across each
domain wall, $\phi$ changes sharply, almost discontinuously on the scale
of the stripes. The direction of the stripes and the domain walls is
defined by the angle $\omega$ with respect to the equator of the
cylinder, illustrated in Fig.~\ref{omegadef}. Let $\theta ^{\prime }$ be
the coordinate running along the $\omega $ direction,
\begin{equation}
\theta ^{\prime }\equiv \theta +\frac zr\tan \omega .
\end{equation}
In terms of this coordinate, our ansatz for the tilt direction $\phi $ can
be written as
\begin{equation}
\label{phi}\phi (\theta ,z)=\phi _0+\frac{n\Delta \phi }{2\pi }\left( \theta
^{\prime }\bmod\frac{2\pi }n\right) ,
\end{equation}
where $\theta ^{\prime }\bmod2\pi /n$ runs from $-\pi /n$ to $\pi /n$. This
ansatz has the sawtooth form shown in Fig.~\ref{sawtooth}. It oscillates
about the average value $\phi _0$ with the amplitude $\Delta \phi $. Thus,
our ansatz has five variational parameters: the radius $r$, the average tilt
direction $\phi _0$, the amplitude of the tilt variation $\Delta \phi $, the
stripe direction $\omega $, and the number of distinct stripes $n$. In terms
of these parameters, the stripe width is
\begin{equation}
L=\left( \frac{2\pi }n\right) r\cos \omega .
\end{equation}

We must now express the free energy~(\ref{free1}) in terms of the five
variational parameters. To simplify the algebra, we suppose that
$\Delta\phi\ll 1$; i.~e., there is only a small variation in the tilt
direction. We will discuss corrections to this approximation at the end
of this section. The curl and divergence of the tilt then become
\begin{equation}
\vec\nabla\times\vec m= \frac{n\Delta\phi\cos\left(\phi_0-\omega\right)}{%
2\pi r\cos\omega}
\end{equation}
and
\begin{equation}
\vec\nabla\cdot\vec m= -\frac{n\Delta\phi\sin\left(\phi_0-\omega\right)}{%
2\pi r\cos\omega}.
\end{equation}
With these expressions, it is straightforward to work out the gradient terms
in Eq.~(\ref{free1}).

The terms $\frac{1}{2}\kappa^{\prime}r^{-2}\cos^2\phi -\lambda_{{\rm HP}%
}r^{-1}\sin\phi\cos\phi$ require special attention. This combination of
terms favors a particular orientation $\phi_0$ of the tilt direction with
respect to the equator of the cylinder. Because it gives an energy penalty
for variations in the tilt away from the optimum direction, it depends on
$\Delta\phi$ as well as on $\phi_0$ and $r$. We expand this combination as a
power series in $(\phi-\phi_0)$, then average it over the cylinder (i.~e.,
average it over the coordinate $\theta^{\prime}$). The result can be written
as $\frac{1}{2}\kappa^{\prime}r^{-2}\cos^2\phi_0 -\lambda_{{\rm HP}%
}r^{-1}\sin\phi_0\cos\phi_0 +\frac{1}{2}\nu(\Delta\phi)^2$, where
\begin{equation}
\label{mudef}\nu=\frac{1}{6}\left[\left(\frac{\kappa^{\prime}}{2r^2}%
\right)^2 +\left(\frac{\lambda_{{\rm HP}}}{r}\right)^2\right]^{1/2}.
\end{equation}
The term $\frac{1}{2}\nu(\Delta\phi)^2$ gives the energy penalty for
variations of $\phi$ away from $\phi_0$.

The domain-wall energy also requires special attention. As noted in Sec.~II,
the continuum free energy of Eqs.~(\ref{freegeneral}) and~(\ref{free1}) does
not describe the narrow regions inside the domain walls. Rather, we must
explicitly add the domain-wall energy $\epsilon_W$ per unit length. In
principle, $\epsilon_W$ can depend on $\phi_0$, $\Delta\phi$, and $\omega$.
In this paper, we will neglect that possible variation and treat $\epsilon_W$
as a constant. In an area $A$ of the membrane, the total length of all the
domain walls is $L_W=A/L$. The domain-wall energy per unit area of membrane
therefore becomes
\begin{equation}
\frac{\epsilon_W L_W}{A}=\frac{\epsilon_W n}{2\pi r\cos\omega}.
\end{equation}

Putting all the pieces together, the total free energy per unit area for our
ansatz now becomes
\begin{eqnarray}
\frac{F_{\rm tot}}{A}=&&\frac{1}{2}\kappa\left(\frac{1}{r}\right)^2
+\frac{1}{2}\kappa' \left(\frac{1}{r}\right)^2 \cos^2 \phi_0\nonumber\\
&&-\lambda_{\rm HP} \left(\frac{1}{r}\right) \sin\phi_0\cos\phi_0
+\frac{1}{2}\nu(\Delta\phi)^2\nonumber\\
&&-\lambda_{\rm LS}
\frac{n\Delta\phi\cos\left(\phi_0-\omega\right)}{2\pi r\cos\omega}
-\frac{\gamma}{r}
\frac{n\Delta\phi\sin\left(\phi_0-\omega\right)}{2\pi r\cos\omega}
\nonumber\\
&&+\frac{1}{2} c \left(\frac{n\Delta\phi}{2\pi r\cos\omega}\right)^2
+\frac{\epsilon_W n}{2\pi r\cos\omega}.
\label{free2}
\end{eqnarray}
This expression must be minimized over the five variational parameters $r$, $%
\phi_0$, $\Delta\phi$, $\omega$, and $n$.

To do this minimization, we will make an approximation: We will treat $n$
as a continuous variable rather than as an integer. This should be a good
approximation for $n\gg1$, although not for $n\approx1$. We will discuss
corrections to this approximation, as well as the previous one, at the
end of this section. After making this approximation, it is convenient to
change variables from $n$ to $L=(2\pi r\cos\omega)/n$. We also change
variables from $\omega$ to $\delta=\phi_0-\omega$. The variable $L$ is
the stripe width, as noted above, and $\delta$ is the difference between
the average tilt direction and the stripe direction. In terms of these
variables, the free energy~(\ref{free2}) simplifies to
\begin{eqnarray}
\frac{F_{\rm tot}}{A}=&&\frac{1}{2}\kappa\left(\frac{1}{r}\right)^2
+\frac{1}{2}\kappa' \left(\frac{1}{r}\right)^2 \cos^2 \phi_0\nonumber\\
&&-\lambda_{\rm HP} \left(\frac{1}{r}\right) \sin\phi_0\cos\phi_0
+\frac{1}{2}\nu(\Delta\phi)^2\nonumber\\
&&-\lambda_{\rm LS}\frac{\Delta\phi\cos\delta}{L}
-\frac{\gamma}{r}\frac{\Delta\phi\sin\delta}{L}\nonumber\\
&&+\frac{1}{2} c \left(\frac{\Delta\phi}{L}\right)^2
+\frac{\epsilon_W}{L}.
\label{free3}
\end{eqnarray}
This expression for the free energy must be minimized over the five
variational parameters $r$, $\phi_0$, $\Delta\phi$, $\delta$, and $L$.

As a first step, we minimize the free energy over the angle $\delta$. We
obtain
\begin{equation}
\label{deltamin}\tan\delta=\frac{\gamma}{r\lambda_{{\rm LS}}}.
\end{equation}
This expression is equivalent to the corresponding result in our earlier
paper~\cite{ss}. This result is important because the angle $\delta$
gives the difference between the stripe direction $\omega$ and the
average tilt direction $\phi_0$. Because these two directions differ by
an angle $\delta$, we can explain the anomaly mentioned at the
end of Sec.~III. Recall that the experiments of
Chung {\it et~al.}~\cite{chung} found one type of tubules with helical
markings at an angle of $11.1^\circ$ from the equator. If these helical
markings indicate the average tilt direction $\phi_0$, then the ratio of
rigidities $(\kappa+\kappa^{\prime})/\kappa$ must be 0.0015, an
anomalously low value. However, if the helical markings indicate the
{\it stripe\/} direction, then this result for
$(\kappa+\kappa^{\prime})/\kappa$ does not hold. That ratio could be
closer to 1, and the angle $\phi_0$ could be closer to $45^\circ$, while
the stripe direction is $11.1^\circ$. Thus, our interpretation is that
those experiments were measuring the stripe direction, not the average
tilt direction.

One might think that $\delta$ would be very small because $\tan\delta$
scales as $1/r$ and $r$ is large. However, in Sec.~III we showed that $r$
scales as $\kappa/\lambda_{{\rm HP}}$. For that reason, $\tan\delta$ scales
as $(\gamma\lambda_{{\rm HP}})/(\lambda_{{\rm LS}}\kappa)$. We have already
argued that the chiral coefficients $\lambda_{{\rm HP}}$ and $\lambda_{{\rm %
LS}}$ should have roughly the same magnitude. Hence, we obtain
\begin{equation}
\tan\delta\sim\frac\gamma\kappa.
\end{equation}
This ratio can be large or small.

For the next step in the calculation, we substitute Eq.~(\ref{deltamin}) for
$\delta$ back into Eq.~(\ref{free3}) for the free energy, to obtain
\begin{eqnarray}
\frac{F_{\rm tot}}{A}=&&\frac{1}{2}\kappa\left(\frac{1}{r}\right)^2
+\frac{1}{2}\kappa' \left(\frac{1}{r}\right)^2 \cos^2 \phi_0\nonumber\\
&&-\lambda_{\rm HP} \left(\frac{1}{r}\right) \sin\phi_0\cos\phi_0
+\frac{1}{2}\nu(\Delta\phi)^2\nonumber\\
&&-\left[\lambda_{\rm LS}^2+\left(\frac{\gamma}{r}\right)^2\right]^{1/2}
\frac{\Delta\phi}{L}\nonumber\\
&&+\frac{1}{2} c \left(\frac{\Delta\phi}{L}\right)^2
+\frac{\epsilon_W}{L}.
\label{free4}
\end{eqnarray}
Minimizing this expression over the stripe width $L$ gives
\begin{equation}
\label{Lmin}L=\frac{c(\Delta\phi)^2}{\left[\lambda_{{\rm LS}%
}^2+(\gamma/r)^2\right]^{1/2} \Delta\phi-\epsilon_W}
\end{equation}
if this denominator is positive, or $L\rightarrow\infty$ otherwise. This
expression for $L$ is the same result that one would expect from studies
of stripes in flat films~\cite{LangerSethna,Hinshaw,jacobs}. It is
equivalent to the result for stripes in tubules in our earlier
paper~\cite{ss}, except that the dependence on $\Delta\phi$ was not
considered there. This expression shows that the $\lambda_{{\rm LS}}$ and
$\gamma/r$ terms in the free energy both tend to induce stripes with a
narrow width. If the energy gain from these terms, added in quadrature,
exceeds the energy cost $\epsilon_W$ of a domain wall, then stripes will
form; otherwise they will not form. If stripes do form, their width
depends on the competition between the Frank free energy proportional to
$c$, which resists variation in $\phi$, and the $\lambda_{{\rm LS}}$ and
$\gamma/r$ terms.

\begin{figure}
\epsfbox{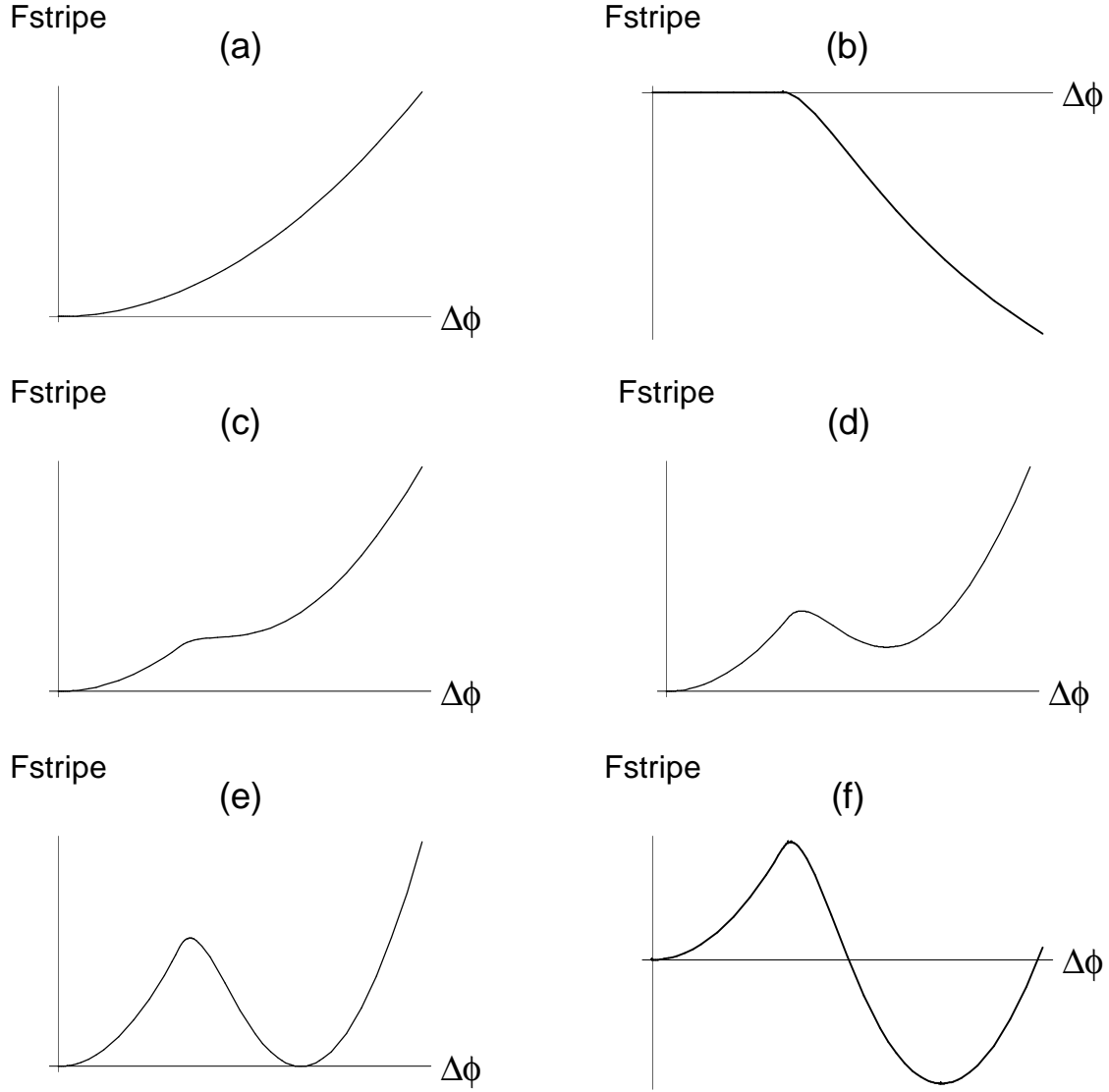}
\caption{Graphical minimization of the stripe free energy
$F_{\rm stripe}$ as a function of $\Delta\phi$, as discussed in the
text.  $F_{\rm stripe}$ is the sum of the two terms plotted in (a) and
(b).  (c)~For large $\nu$, the only minimum is the uniform state with
$\Delta\phi=0$.  (d)~As $\nu$ decreases, another minimum develops for
$\Delta\phi\neq0$.  (e)~At $\nu=\nu_c$, there is a first-order
transition from the uniform state ($\Delta\phi=0$) to the modulated state
($\Delta\phi=\Delta\phi_c$).  (f)~For $\nu<\nu_c$, the modulated state
becomes the absolute minimum of the free energy.}
\label{graphicalmin}
\end{figure}

We now insert Eq.~(\ref{Lmin}) for $L$ back into Eq.~(\ref{free4}) for the
free energy. The result is
\begin{eqnarray}
\frac{F_{\rm tot}}{A}=&&\frac{1}{2}\kappa\left(\frac{1}{r}\right)^2
+\frac{1}{2}\kappa' \left(\frac{1}{r}\right)^2 \cos^2 \phi_0\nonumber\\
&&-\lambda_{\rm HP} \left(\frac{1}{r}\right) \sin\phi_0\cos\phi_0
+\frac{1}{2}\nu(\Delta\phi)^2\nonumber\\
&&-\frac{1}{2c}\left[
\left[\lambda_{\rm LS}^2+\left(\frac{\gamma}{r}\right)^2\right]^{1/2}
-\frac{\epsilon_W}{\Delta\phi}\right]^2.
\label{free5}
\end{eqnarray}
if the quantity in brackets in the last term is positive. If not, then the
last term is zero. Because only the last two terms in the free energy depend
on $\Delta \phi $, we will call them $F_{{\rm stripe}}(\Delta \phi )/A$. We
must now minimize $F_{{\rm stripe}}$ over $\Delta \phi $. This minimization
cannot be done analytically, but it can be done graphically. In Fig.~\ref
{graphicalmin}, we plot $F_{{\rm stripe}}$ as a function of $\Delta \phi $
for a sequence of values of $\nu $. The parameter $\nu $, defined in Eq.~(%
\ref{mudef}), gives the energy penalty for variations of $\phi $ away from $%
\phi _0$. For large $\nu $, the only minimum of $F_{{\rm stripe}}$ is for $%
\Delta \phi =0$. That is reasonable, because a large value of $\nu $
suppresses all variations in $\phi $. As $\nu $ decreases, $F_{{\rm stripe}}$
develops a local minimum for $\Delta \phi \neq 0$. At a critical value $\nu
_c$, there is a first-order transition from $\Delta \phi =0$ to $\Delta \phi
=\Delta \phi _c$. For $\nu <\nu _c$, the minimum at $\Delta \phi \neq 0$
becomes the absolute minimum of the free energy. To calculate $\nu _c$ and $%
\Delta \phi _c$, we set $F_{{\rm stripe}}=0$ and $\partial F_{{\rm stripe}%
}/\partial (\Delta \phi )=0$. The result is
\begin{equation}
\nu _c=\frac 1{16c\epsilon _W^2}\left[ \lambda _{{\rm LS}}^2+\left( \frac
\gamma r\right) ^2\right] ^2
\end{equation}
and
\begin{equation}
\Delta \phi _c=\frac{2\epsilon _W}{\left[ \lambda _{{\rm LS}}^2+(\gamma
/r)^2\right] ^{1/2}}.
\end{equation}
The corresponding value of the stripe width is
\begin{equation}
L_c=\frac{4c\epsilon _W}{\lambda _{{\rm LS}}^2+(\gamma /r)^2},
\end{equation}
and hence the gradient in the tilt direction $\phi $ is
\begin{equation}
\label{DphiOverL}\frac{\Delta \phi _c}{L_c}=\frac{\left[ \lambda _{{\rm LS}%
}^2+(\gamma /r)^2\right] ^{1/2}}{2c}.
\end{equation}

In summary, for $\nu>\nu_c$ the tubules are uniform, with no tilt
modulation. In that case, we have $\Delta\phi=0$, $L\rightarrow\infty$, and $%
F_{{\rm stripe}}=0$. At $\nu=\nu_c$, there is a first-order transition from
the uniform state to a modulated state with $\Delta\phi=\Delta\phi_c$ and $%
L=L_c$. At this point, $F_{{\rm stripe}}$ is still zero. For $\nu<\nu_c$,
the tubules remain in the modulated state. As $\nu$ decreases, $\Delta\phi$
and $L$ both grow larger and $F_{{\rm stripe}}$ becomes negative. It is not
possible to derive a general analytic expression for $\Delta\phi$ and $L$ in
the modulated state. However, a good estimate would be $\Delta\phi\approx%
\Delta\phi_c$ and $L\approx L_c$.

We must now minimize the free energy~(\ref{free5}) over the radius $r$ and
the average tilt direction $\phi_0$, which are the remaining variational
parameters. For $\nu\geq\nu_c$, we have $F_{{\rm stripe}}=0$, and hence the
total free energy is
\begin{eqnarray}
\frac{F_{\rm tot}}{A}=&&\frac{1}{2}\kappa\left(\frac{1}{r}\right)^2
+\frac{1}{2}\kappa' \left(\frac{1}{r}\right)^2 \cos^2 \phi_0\nonumber\\
&&-\lambda_{\rm HP} \left(\frac{1}{r}\right) \sin\phi_0\cos\phi_0.
\end{eqnarray}
This is exactly the free energy considered in Sec.~III. Hence, the solution
given in Eqs.~(\ref{rmin}) and~(\ref{phi0min}) still applies to the uniform
state of tubules. For $\nu<\nu_c$, we have $F_{{\rm stripe}}<0$ and hence
\begin{eqnarray}
\frac{F_{\rm tot}}{A}=&&\frac{1}{2}\kappa\left(\frac{1}{r}\right)^2
+\frac{1}{2}\kappa' \left(\frac{1}{r}\right)^2 \cos^2 \phi_0\nonumber\\
&&-\lambda_{\rm HP} \left(\frac{1}{r}\right) \sin\phi_0\cos\phi_0
+\frac{F_{\rm stripe}}{A}.
\end{eqnarray}
Although we do not have a general expression for $F_{{\rm stripe}}$, we see
that $F_{{\rm stripe}}$ can be expanded in a power series in $r^{-2}$. The
constant term in that series does not affect the minimization. The term
proportional to $r^{-2}$ just gives a {\it negative renormalization\/} of
the rigidity $\kappa$. The higher-order terms in $r^{-2}$ should be
neglected, because we have already neglected terms of that order in our
original curvature expansion for the free energy. For that reason, if we
just use the renormalized rigidity $\kappa_R$ in place of the bare rigidity $%
\kappa$, then the derivation of $r$ and $\phi_0$ in Sec.~III still follows.
We obtain Eqs.~(\ref{rmin}) and~(\ref{phi0min}) for $r$ and $\phi_0$ with
the renormalized $\kappa_R$ in place of $\kappa$.

At this point, we have minimized the free energy~(\ref{free3}) over the five
parameters in our variational ansatz: $r$, $\phi_0$, $\Delta\phi$, $\delta$,
and $L$. We have shown that there is a first-order transition from a uniform
state to a modulated state characterized by stripes in the tilt direction.
We must now discuss corrections to two approximations in this variational
calculation.

First, we assumed that there is only a small variation in the tilt
direction across a stripe:  $\Delta\phi\ll 1$. In some systems,
$\Delta\phi$ might be locked at a larger angle. For example, in a
membrane with hexatic bond-orientational order, the angle
$\Delta\phi=60^\circ$ would be favored, because that angle would allow
the tilt direction to jump from one bond direction to another bond
direction across each domain wall. Such locking has been investigated in
the context of flat liquid-crystal films by
Hinshaw {\it et al.}~\cite{Hinshaw}. In such a system, the transition
from the uniform state to the modulated state would become more strongly
first-order than we calculated above:  the larger discontinuity in
$\Delta\phi$ would imply a larger latent heat of transition. Thus, the
quantitative predictions of this section would no longer apply.
Nevertheless, our qualitative predictions would still apply, because
tubules would have a first-order transition from a uniform state to a
modulated state whether or not $\Delta\phi\ll 1$.

Second, we did the variational calculation with the approximation that
the stripe width $L$ is a continuous variable, or equivalently, that the
number of stripes $n$ takes continuous rather than just integer values.
In fact, because of the periodic boundary conditions going around a
cylinder, the parameter $n$ must be an integer, and hence
$L=(2\pi r\cos\omega)/n$ is restricted to a discrete set of values. For
that reason, there will be a series of jumps, or first-order transitions,
between different integer values of $n$. Within each constant-$n$
``phase,'' there will be one special point where the continuous
minimization gives exactly the right answer: $n=$~integer. At those
special points, all the results of the minimization (for $\delta$, $L$,
$\Delta\phi$, etc.) will be correct. Between those special points, there
will be deviations from the predictions of the continuous approximation.
The actual values should therefore fluctuate {\it about\/} the values
from the continuous approximation. Thus, this approximation should give a
reasonable estimate of the actual results.

\section{Helical Ribbons}

In this section, we apply the theory for the modulated state of tubules,
developed in the previous section, to helical ribbons. We draw an
explicit analogy between a single stripe of a modulated tubule and a
single ribbon. We show that a helical ribbon can be either a stable state
of a membrane or an unstable intermediate state in the formation of a
tubule. A stable helical ribbon has a particular optimum width. If more
molecules are added to the ribbon, it becomes longer, not wider. By
contrast, an unstable helical ribbon grows wider until it forms a tubule.

\begin{figure}
\epsfbox{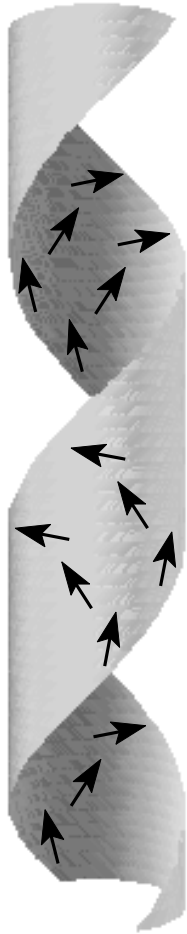}
\caption{Schematic view of a helical ribbon, with a variation in the tilt
direction across the width of the ribbon.  The arrows represent the tilt
direction on one side of the bilayer.}
\label{ribbongeometry}
\end{figure}

The basis of our calculation is the geometry shown in Fig.~\ref
{ribbongeometry}. A narrow ribbon winds helically around the $z$ axis. We
consider a linear variation of the tilt direction $\phi $ up to the ribbon
edge. The tilt variation is in the direction specified by the angle $\omega $
with respect to the equator, as is the ribbon itself. Again, we describe the
tilt variation using the $\theta ^{\prime }$ coordinate,
\begin{equation}
\label{ThetaPrime}\theta ^{\prime }\equiv \theta +\frac zr\tan \omega .
\end{equation}
which runs in the $\omega $ direction. The ribbon edges are at $\theta
^{\prime }=\pm \theta ^{*}$, where $\theta ^{*}$ is a variational parameter.
The maximum possible value is $\theta ^{*}=\pi $, at which point the edges
of the ribbon collide, and the ribbon forms a tubule. Between the edges, our
ansatz for the tilt direction $\phi $ is
\begin{equation}
\phi (\theta ,z)=\phi _0+\frac{\Delta \phi }{2\theta ^{*}}\theta ^{\prime }.
\end{equation}
This ansatz is a single period of the sawtooth wave of Fig.~\ref{sawtooth}.
As in Sec.~IV, the ansatz has five variational parameters: the radius $r$,
the average tilt direction $\phi _0$, the amplitude of the tilt variation $%
\Delta \phi $, the ribbon direction $\omega $, and the angle $\theta ^{*}$,
which is the half-width of the ribbon in the $\theta ^{\prime }$ coordinate.
The actual width of the ribbon is
\begin{equation}
L=2\theta ^{*}r\cos \omega .
\end{equation}

We must now express the free energy~(\ref{free1}) of a helical ribbon in
terms of these five variational parameters. The free energy is expressed per
unit area of membrane, where the area does not include the gaps between the
edges of the ribbon. For most of the terms, the free energy per unit area of
a tubule, derived in Sec.~IV, still applies. However, instead of the
domain-wall energy $\epsilon_W$, we must now use the edge energy $\epsilon_E$
per unit length for each of the two edges of the ribbon. This edge energy
gives the energy cost of having different parts of the molecules exposed to
the solvent at the ribbon edges. Hence, the total free energy per unit area
in our ansatz for a helical ribbon becomes
\begin{eqnarray}
\frac{F_{\rm tot}}{A}=&&\frac{1}{2}\kappa\left(\frac{1}{r}\right)^2
+\frac{1}{2}\kappa' \left(\frac{1}{r}\right)^2 \cos^2 \phi_0\nonumber\\
&&-\lambda_{\rm HP} \left(\frac{1}{r}\right) \sin\phi_0\cos\phi_0
+\frac{1}{2}\nu(\Delta\phi)^2\nonumber\\
&&-\lambda_{\rm LS}
\frac{\Delta\phi\cos\left(\phi_0-\omega\right)}{2\theta^* r\cos\omega}
-\frac{\gamma}{r}
\frac{\Delta\phi\sin\left(\phi_0-\omega\right)}{2\theta^* r\cos\omega}
\nonumber\\
&&+\frac{1}{2} c \left(\frac{\Delta\phi}{2\theta^* r\cos\omega}\right)^2
+2\cdot\frac{\epsilon_E}{2\theta^* r\cos\omega}.
\label{free6}
\end{eqnarray}
This expression must be minimized over the five variational parameters $r$, $%
\phi_0$, $\Delta\phi$, $\omega$, and $\theta^*$.

At this point, we note that all five variational parameters are continuous
variables; none of them is an integer, as in the previous section. Thus, we
can immediately change variables from $\theta^*$ to the ribbon width $%
L=2\theta^* r\cos\omega$, and from $\omega$ to the difference angle $%
\delta=\phi_0-\omega$. In terms of these variables, the ribbon free energy~(%
\ref{free6}) simplifies to
\begin{eqnarray}
\frac{F_{\rm tot}}{A}=&&\frac{1}{2}\kappa\left(\frac{1}{r}\right)^2
+\frac{1}{2}\kappa' \left(\frac{1}{r}\right)^2 \cos^2 \phi_0\nonumber\\
&&-\lambda_{\rm HP} \left(\frac{1}{r}\right) \sin\phi_0\cos\phi_0
+\frac{1}{2}\nu(\Delta\phi)^2\nonumber\\
&&-\lambda_{\rm LS}\frac{\Delta\phi\cos\delta}{L}
-\frac{\gamma}{r}\frac{\Delta\phi\sin\delta}{L}\nonumber\\
&&+\frac{1}{2} c \left(\frac{\Delta\phi}{L}\right)^2
+\frac{2\epsilon_E}{L}.
\label{free7}
\end{eqnarray}
Equation~(\ref{free7}) for the ribbon free energy is exactly the same as
Eq.~(\ref{free3}) for the tubule free energy, but with $2\epsilon_E$
substituted in place of $\epsilon_W$. However, for ribbons, all of
the parameters really are continuous variables; that is not just an
approximation. Thus, the theory of ribbons is mathematically simpler than
the theory of tubules.

Because the free energy for ribbons is equivalent to the free energy for
tubules, we can carry over our results from the theory of tubules in
Sec.~IV. The direction of a ribbon, relative to the average tilt
direction, is given by the difference angle
\begin{equation}
\label{Tandelta}\tan \delta =\frac \gamma {r\lambda _{{\rm LS}}}\sim \frac
\gamma \kappa .
\end{equation}
The ribbon width $L$ is given by
\begin{equation}
L=\frac{c(\Delta \phi )^2}{\left[ \lambda _{{\rm LS}}^2+(\gamma /r)^2\right]
^{1/2}\Delta \phi -2\epsilon _E}
\end{equation}
if this denominator is positive, or $L\rightarrow \infty $ otherwise.
However, note that the maximum possible value of $L$ is $L_{{\rm max}}=2\pi
r\cos \omega $. If $L>L_{{\rm max}}$, the edges collide and the ribbon forms
a tubule. Thus, there is a stable ribbon with $L<L_{{\rm max}}$ only for a
certain window of parameters.

For the amplitude $\Delta \phi $ of the tilt variation, the analysis of
Sec.~IV applies again. At a critical value of $\nu $,
\begin{equation}
\nu _c=\frac 1{64c\epsilon _E^2}\left[ \lambda _{{\rm LS}}^2+\left( \frac
\gamma r\right) ^2\right] ^2,
\end{equation}
there is a first-order transition from a state with $\Delta \phi =0$ and $%
L\rightarrow \infty $, i.~e., a state with no stable ribbon, to a state with
$\Delta \phi =\Delta \phi _c$ and $L=L_c$, where
\begin{equation}
\Delta \phi _c=\frac{4\epsilon _E}{\left[ \lambda _{{\rm LS}}^2+(\gamma
/r)^2\right] ^{1/2}}
\end{equation}
and
\begin{equation}
L_c=\frac{8c\epsilon _E}{\lambda _{{\rm LS}}^2+(\gamma /r)^2}.
\end{equation}
If $L_c<L_{{\rm max}}$, then a stable ribbon forms at this transition.
Finally, the minimization over $r$ and $\phi _0$ goes through exactly as in
the case of the modulated state of tubules. The results are the same as in
Sec.~III, but with the renormalized rigidity $\kappa _R$ in place of the
bare rigidity $\kappa .$

\section{Ripples}

In this section, we consider the possibility of non-uniformly curved
states of a tubule. In particular, we consider equilibrium undulated, or
rippled, textures on otherwise cylindrical tubules. In part, this is
because of the similarity of the model described above in
Eq.~(\ref{freegeneral}) to recent models of modulated phases of lipid
bilayers~\cite{Pbeta1,Pbeta2} and liquid crystal films~\cite{Spt}, and
the suggestion in Ref.~\cite {Pbeta2} that chiral stripe textures such as
those of Sec.~IV may exhibit a rippled shape. There have also been
unpublished experimental reports of regular periodic undulations of the
surface of tubules~\cite{NoteBritt}. Quite generally, however, we expect
that an underlying chiral stripe texture, with a spatially varying tilt
field as described in Sec.~IV, will lead to a shape modulation of the
tubule, as shown in Fig.~\ref{ripple3D}. This is because of the coupling
of the molecular orientation, or tilt field, to membrane
shape~\cite{Pbeta1}.

Given an underlying stripe texture of the molecular tilt $\vec m(\theta ,z)$%
, we study the corresponding tubule shape that results from the coupling of
molecular tilt to membrane shape. Among the possible explicit couplings of
molecular tilt to membrane shape, the term $\gamma K_a^aD_bm^b$ is general
to all membranes with in-plane orientational order. It is allowed by
symmetry for both chiral and achiral lipid bilayers. Furthermore, the
coupling constant $\gamma $ is expected to be of the same order as the
bending modulus $\kappa$~\cite{Shapes}. The second coupling in increasing
powers of $m$ and $K$, $\lambda _{{\rm HP}}\sqrt{g}\epsilon _{bc}m^am^cK_a^b$%
, is permitted only for chiral systems. Here, we shall consider a somewhat
simplified model, in which we let $\kappa ^{\prime }=0$ in Eq.~(\ref
{freegeneral}). We calculate the modulated shapes of tubules with tilt
modulation in the limit that the ripple amplitude is small compared with the
tubule radius $r$, which is of order 1~$\mu $m.

For a rippled surface, of course, the curvature tensor $K_{ab}$ in Eq.~(\ref
{freegeneral}) is no longer constant. In Sec.~IV we described the tilt field
$\vec m(\theta ,z)$ by a tilt angle $\phi (\theta ,z)$. Here, we consider
also a ripple characterized by a deviation $h(\theta ,z)$ of the membrane
surface away from a background cylindrical geometry. More precisely, the
membrane position in three dimensions is given by
\begin{equation}
\vec X(\theta ,z)=\left( (r+h)\cos \theta ,(r+h)\sin \theta ,z\right) ,
\end{equation}
where $r$ is the fixed (average) radius of the tubule. In other words, we
consider a small undulation $h$ of the local cylinder radius. The following
analysis is valid for $h\ll r$.

We find it convenient to use coordinates $(s,z)$, where $s=r\theta $. In
these coordinates, the metric tensor for the tubule surface is
\begin{equation}
g_{ab}=\vec t_a\cdot \vec t_b=\left(
\begin{array}{cc}
(1+h/r)^2+(\partial _sh)^2 & \partial _sh\partial _zh \\
\partial _sh\partial _zh & 1+(\partial _zh)^2
\end{array}
\right) ,
\end{equation}
where $\vec t_a=\partial _a\vec X$ form a basis for the local tangent plane
to the membrane. Through second order in the, presumed small, height
modulation $h$, the surface area measure is $dA=\sqrt{g}dsdz$, where
\begin{equation}
\sqrt{g}=\sqrt{\left| \det g_{ab}\right| }\cong \left( 1+\frac hr\right)
\left( 1+{\frac 12}(\vec \nabla h)^2\right) ,
\end{equation}
and
\begin{equation}
(\vec \nabla h)^2=(\partial _sh)^2+(\partial _zh)^2.
\end{equation}
The curvature tensor is $K_{ab}=\hat N\cdot \partial _a\partial _b\vec X$,
where $\hat N=\vec t_\theta \times \vec t_z/|\vec t_\theta \times \vec t_z|$
is the surface normal. Through order $h^2$, the mean curvature
\begin{equation}
\label{TraceK}K_a^a=g^{ab}K_{ab}\cong -\frac 1r+\nabla ^2h+{\frac{(\vec
\nabla h)^2}{2r}}-\frac{\left( \partial _sh\right) ^2}r,
\end{equation}
where $g^{ab}=\left( g_{ab}\right) ^{-1}$, and
\begin{equation}
\nabla ^2h=\partial _s^2h+\partial _z^2h.
\end{equation}
We shall find that the tilt modulation described in Sec. IV leads to a
modulation of $h$ with the same period. For the stripe texture
considered above, this period is the stripe width $L$. The last two terms in
Eq.~(\ref{TraceK}) are of order $h^2/\left( rL^2\right) $, which is smaller
than $\nabla ^2h\sim h/L^2$ by approximately $h/r$. Thus, we shall retain
only the first two terms in Eq.~(\ref{TraceK}).

To order $h$, the divergence of $\vec m$ is given by
\begin{equation}
\label{Divm}D_am^a\cong \left( 1-\frac hr\right) \partial _s\cos \phi
+\partial _z\sin \phi +\frac{\partial _zh}r\sin \phi .
\end{equation}
The first terms in Eq.~(\ref{Divm}) are of order $\Delta \phi /L$. From Eqs.
(\ref{DphiOverL}) and (\ref{Tandelta}), we expect that this is of order
\begin{equation}
\label{LambdaLSRelation}\frac{\Delta \phi }L\sim \frac{\lambda _{{\rm LS}}}c%
\sim \frac 1r
\end{equation}
near the transition to the stripe texture, since we also expect that $\kappa
$ and $c$ are of the same order~\cite{Shapes}. The last term in Eq.~(\ref
{Divm}) is thus smaller than the other terms by a factor of approximately $%
h/L$. Below, we shall ignore the last term in Eq.~(\ref{Divm}). We comment
on the validity of this below.

The leading-order terms in Eq.~(\ref{freegeneral}) that depend on the
ripple shape $h$ are given by
\begin{eqnarray}
\label{Fh}
F_h = \int
&& \sqrt{g}dsdz\left[ {\frac \kappa 2}\left( K_a^a\right) ^2+\gamma \left(
K_a^a\right) \left( D_bm^b\right) \right] \nonumber\\  \simeq \int && dsdz
\Biggl[\frac \kappa 2\left( \frac 1{r^2}-2\frac{\nabla ^2h}r+\left( \nabla
^2h\right) ^2\right)                  \nonumber\\  &&
-\gamma \left( \nabla ^2h-\frac 1r\right)
\left( \partial _s\cos \phi +\partial _z\sin \phi \right) \nonumber\\ &&
+\lambda_{\rm HP}\left(\sin\phi\cos\phi
     \left(\partial^2_sh-\partial^2_zh\right)
    +\left(\sin^2\phi-\cos^2\phi\right)\partial_s\partial_zh\right)\Biggr]
\end{eqnarray}
For a fixed background cylinder radius $r$ and a (predetermined) modulating
tilt angle $\phi (\theta ,z)$, which following Sec. IV can be described in
terms of a single variable $\theta ^{\prime }$ given in Eq.~(\ref{ThetaPrime}%
), the Euler-Lagrange equation for the height field $h$ is determined by
variation of Eq.~(\ref{Fh}) with respect to $h$. To leading order in $h/r$,
the result is an ordinary differential equation,
\begin{equation}
\kappa \partial _{s^{\prime }}^4h-\gamma \partial _{s^{\prime }}^3\cos
\left( \phi -\omega \right) +\frac{\lambda _{{\rm HP}}}2\partial _{s^{\prime
}}^2\sin 2\left( \phi -\omega \right) =0,
\end{equation}
that can be integrated to yield
\begin{equation}
\label{Soln}\kappa \partial _{s^{\prime }}^2h-\gamma \partial _{s^{\prime
}}\cos \left( \phi -\omega \right) +\frac{\lambda _{{\rm HP}}}2\sin 2\left(
\phi -\omega \right) =\text{constant},
\end{equation}
where $s^{\prime }=r\cos \omega \theta ^{\prime }=s\cos \omega +z\sin \omega
$ extends from $-L/2$ to $L/2$. Here, we have also used the fact that $h$ is
a periodic function. This does not allow, for instance, a linear term in
Eq.~(\ref{Soln}). For the tilt field given above in Eq.~(\ref{phi}),
\begin{equation}
\label{ELeq}\partial _{s^{\prime }}\cos \left( \phi -\omega \right) =-\frac{%
\Delta \phi }L\sin \left( \phi -\omega \right) .
\end{equation}
Within one period of the modulated texture, the tilt angle $\phi $ can be
represented as
\begin{equation}
\phi -\omega =\delta +\Delta \phi \frac n{2\pi L}s^{\prime },
\end{equation}
where again $-L/2<s^{\prime }<L/2$. For small $\Delta \phi $, the
solutions to Eq.~(\ref{ELeq}) can be written
\begin{equation}
\label{GenSoln}h\cong h_1\left( \frac{\left( s^{\prime }\right) ^2}2-\alpha
\right) +\frac{\Delta \phi }Lh_2\left( \frac{\left( s^{\prime }\right) ^3}6%
-\beta s^{\prime }\right) ,
\end{equation}
where $\alpha $ and $\beta $ are constants, and the coefficients $h_1$ and $%
h_2$ depend on $\gamma $, $\kappa $, $\lambda _{{\rm HP}}$, $\delta $, $%
\Delta \phi $, and $L$.

At this point, an additional assumption concerning the domain wall is
necessary. Here we consider two possibilities, both of which, however, yield
the same functional form (Eq.~[\ref{GenSoln}]) for the height ripples on the
surface of tubules. As noted above, we have not taken account of possible
dependencies of the domain wall energy $\epsilon _W$ on parameters of our
model such as $\phi $ or $\omega $. If we continue to assume as before that
the domain wall is of infinitesimal thickness, and costs an energy $\epsilon
_W$ per unit length that is independent of model parameters, which now
include the possibility of a finite slope discontinuity of the membrane at
the domain wall, then the constant in Eq.~(\ref{Soln}) can be determined by
minimizing the integrated free energy of Eq.~(\ref{Fh}) using Eq.~(\ref{Soln}%
). Because of the presence of the domain wall, we must include total
derivatives such as the term $\kappa \left( \nabla ^2h\right) /r$ in Eq.~(%
\ref{Fh}). This is equivalent to solving Eq.~(\ref{Soln}) with the constant
on the right hand side set equal to $1/r$. The result is given by Eq.~(\ref
{GenSoln}) with
\begin{equation}
h_1=\left( \frac 1r-\frac \gamma \kappa \frac{\Delta \phi }L\sin \delta -%
\frac{\lambda _{{\rm HP}}}{2\kappa }\sin 2\delta \right)
\end{equation}
and
\begin{equation}
\label{h2}h_2=-\left( \frac \gamma {6\kappa }\frac{\Delta \phi }L\cos \delta
+\frac{\lambda _{{\rm HP}}}{6\kappa }\cos 2\delta \right) .
\end{equation}
For the above to be a periodic function in the range
from $-L/2$ to $L/2$, we must have
\begin{equation}
\label{beta}\beta =\frac{L^2}{24}.
\end{equation}
The other constant $\alpha $ is arbitrary. We note that in the above, the
contributions to both $h_1$ and $h_2$ from the chiral and achiral
couplings are of the same order, provided that $\lambda _{{\rm HP}}\approx
\lambda _{{\rm LS}}$ and $\gamma \approx \kappa $, since $\Delta \phi /L\sim
1/r$.

On the other hand, if we assume a narrow but finite domain wall region (2),
in which the elastic constants $\kappa $, $\gamma $, and $\lambda _{{\rm HP}%
} $ have the same values as in the region (1) above, and in which the tilt
angle $\phi $ varies again linearly, but in reverse to its variation in
region (1), then the solution in region (1) can still be expressed by Eq.~(%
\ref{GenSoln}), where
\begin{equation}
h_1=-\frac \gamma \kappa \frac{\Delta \phi }L\sin \delta ,
\end{equation}
$h_2$ is given by Eq.~(\ref{h2}), and $\beta $ is given by Eq.~(\ref{beta}).
This is valid for domain walls of width $L^{\prime }\ll L$. The general
result for $L^{\prime }\approx L$ is somewhat more complicated, although the
general form of Eq.~(\ref{GenSoln}) is still valid. In particular,
periodicity of $h$ in region (1) alone is no longer valid, and hence, the
coefficient $\beta $ differs from the value above. (A somewhat more general
solution was derived in Ref.~\cite{Pbeta2} for flat membranes.) In general,
however, the membrane slope is continuous across the boundaries between
regions (1) and (2) under the conditions of equal elastic constants $\kappa $%
, $\gamma $, and $\lambda _{{\rm HP}}$. We note, however, that the chiral
and achiral couplings no longer contribute to the ripple amplitude at the
same order. The dominant contribution to the ripple amplitude is from the
achiral term $\gamma K_a^aD_bm^b$, while the contribution to Eq.~(\ref
{GenSoln}) from the chiral coupling $\lambda _{{\rm HP}}\sqrt{g}\epsilon
_{bc}m^am^cK_a^b$ is smaller by a factor of order $\Delta \phi $.

For a stripe width $L$, the amplitude of the height modulation $h$ is of
order $L^2/r$, where we have used Eq.~(\ref{LambdaLSRelation}). Thus,
\begin{equation}
\frac hr\sim \left( \frac Lr\right) ^2\sim \frac 1{n^2},
\end{equation}
where, as in Sec. IV, $n$ is the number of stripes on the cylinder. We have
assumed that this is small. So, our analysis above is valid for $L\ll r$. In
other words, we have calculated the shape of tubules with stripe textures in
the limit that the stripe texture has both amplitude and period small
compared with the tubule radius. Note also that $h$ is smaller than $L$ by a
factor of $L/r\sim 1/n$.

\begin{figure}
\centering\leavevmode\epsfysize=8.3in\epsfbox{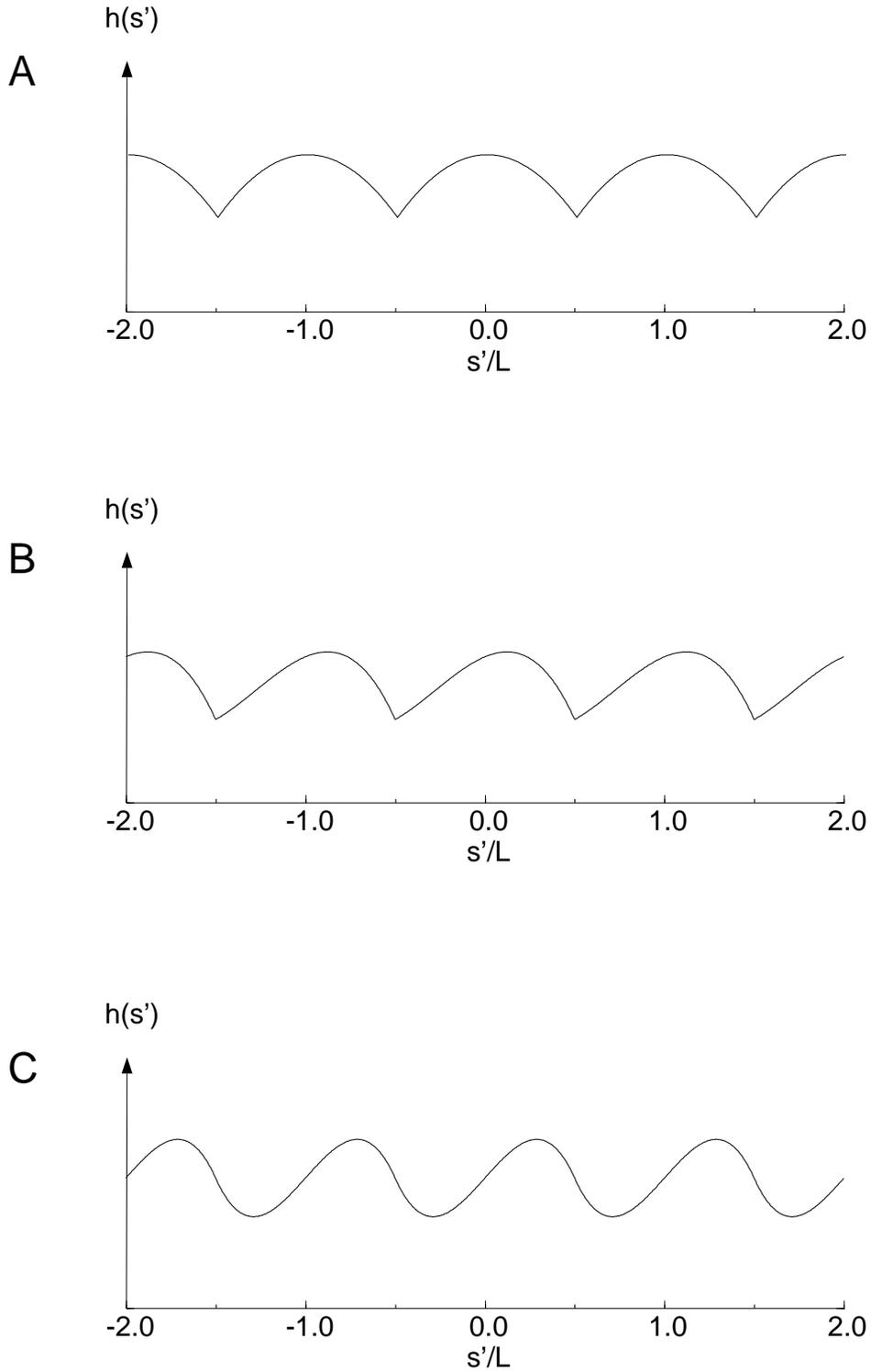}
\caption{Plots of the ripple shapes, as a function of the coordinate $s'$
normal to the ripples.  For small $\Delta\phi$, the dominant contribution
is symmetric under $s'\rightarrow-s'$, but corrections are asymmetric.
(a)~$h_1<0$ and $h_2=0$.  (b)~$h_2\Delta\phi=0.3h_1.$
(c)~$h_2\Delta\phi=h_1.$}
\label{rippleshapes}
\end{figure}

In Fig.~\ref{rippleshapes}, we show representative ripple shapes for
various values of $h_1$ and $h_2$. In Fig.~\ref{rippleshapes}a, we show
the shape for $h_1<0$ and $h_2=0$. This is the dominant contribution for
small $\Delta \phi $. This term is symmetric under
$s^{\prime}\rightarrow -s^{\prime }$. The correction to this, smaller by
order $\Delta \phi $, is asymmetric. In Figs.~\ref{rippleshapes}b and
\ref{rippleshapes}c, we show the corresponding shapes for
$h_2\Delta\phi=0.3h_1$ and $h_2\Delta\phi=h_1$.

As a final point, note that the analysis in this section, as well as
Sec.~IV, applies to any membrane in a cylindrical morphology, regardless
of how it formed.  In particular, if any membrane is adsorbed onto a
pre-existing cylindrical substrate (perhaps a microscopic wire or fiber),
it can form stripes in the tilt direction and ripples in the curvature.
These modulations can occur even if the membrane is not chiral:  In a
non-chiral membrane, they would be induced by the $\gamma$ term in the
free energy, which couples membrane curvature to variations in the
direction of the tilt.  These modulations can reduce the free energy by
concentrating the curvature into domain walls.

\section{Discussion}

In this paper, we have presented a general theory of tubules and helical
ribbons based on the concept of chiral molecular packing. This theory
shows that tubules can have both uniform and modulated states. In the
uniform state, tubules have a constant orientation of the molecular tilt
with respect to the equator of the cylinder. In the modulated state,
tubules have a periodic, helical modulation in the direction of the
molecular tilt, and corresponding ripples in the curvature of the
cylinders. In this section, we discuss the experimental evidence for
these theoretical predictions.

There are two types of experimental evidence supporting the concept that
the formation of tubules and helical ribbons is due to chiral molecular
packing. First, many experiments have seen helical markings that wind
around tubules, giving tubules a chiral substructure. Clearly, helical
ribbons always have a chiral structure. It is reasonable that the
observed chirality of these microstructures results from a chiral packing
of the molecules. Second, recent experiments have found that diacetylenic
lipid tubules have a very strong circular dichroism, which indicates a
local chiral packing of the molecules, regardless of whether a chiral
pattern is visible on the surface of the cylinder~\cite{cd}. The same
diacetylenic lipid molecules in solution or in large spherical vesicles
have very low circular dichroism. These results show that the molecular
packing in tubules is chiral, while the molecular packing in spherical
vesicles is not chiral.

\begin{figure}
\centering\leavevmode\epsfysize=8.3in\epsfbox{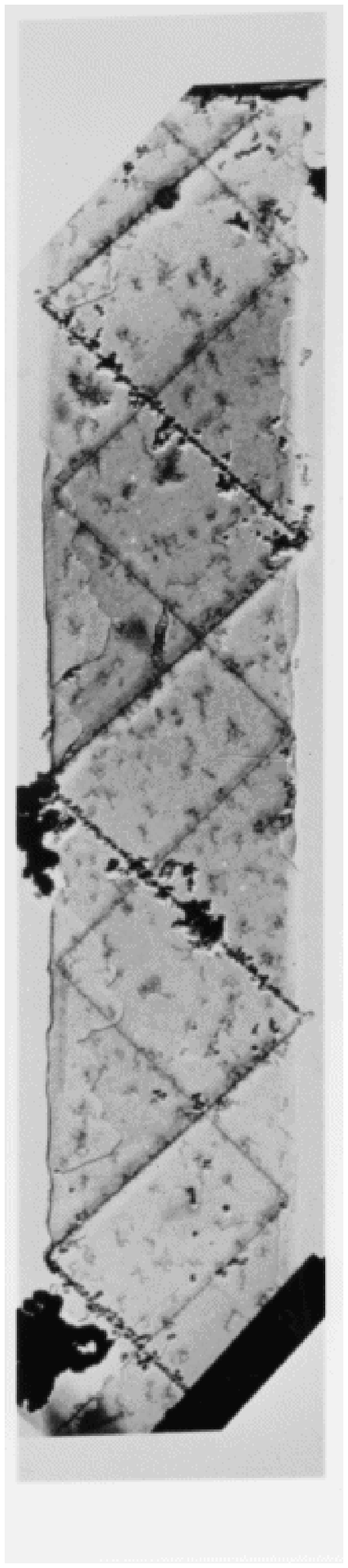}
\caption{Transmission electron micrograph of a tubule with adsorbed Pd/Ni
catalyst particles on the surface~\protect\cite{gulik}.  The diameter of
the tubule is approximately 0.5~$\mu$m.  Reprinted from
Ref.~\protect\cite{ss}.}
\label{tubulephoto}
\end{figure}

So far, there has not been any direct test of our prediction of tilt
modulation---no experiments have been sensitive to the local direction of
molecular tilt in tubules. However, the helical markings on tubules
provide indirect evidence for this prediction. In some cases, these
helical markings are boundaries between sections of tubules with
different numbers of bilayers in the walls. However, in other cases,
helical markings appear even when there is no detectable discontinuity in
the number of bilayers, as in Fig.~\ref{tubulephoto}. These helical
markings are apparently stable, because they do not anneal away in time,
and hence seem to be a characteristic of the equilibrium state of
tubules. Our interpretation is that these helical markings are the
orientational domain walls predicted by our theory. In this
interpretation, the domain walls are visible in electron micrographs
because impurities accumulate there and colloidal particles from the
solution preferentially adsorb there. Such preferential diffusion of
impurities to orientational domain walls has been observed directly in
Langmuir monolayers~\cite{qiu}.

Of course, this interpretation of the observed helical markings provides
only indirect support for our theory. For a more direct test of our
theory, one would need an experiment that can directly probe the local
direction of molecular tilt. One possible experimental technique is
fluourescence microscopy with polarized laser excitation. This technique
has been used to observe variations in the local tilt direction in
Langmuir monolayers~\cite{qiu}. To apply this technique to tubules, one
would either (a)~use the intrinsic fluorescence of the constituent
amphiphilic molecules, (b)~attach a fluorescent group to the molecules,
or (c)~put a fluorescent probe into the membrane that forms
tubules. One would then illuminate the tubules with polarized light
from a laser source. Variations in the direction of molecular tilt would
then lead to variations in the intensity of fluorescence, which could be
detected using confocal microscopy or near-field scanning optical
microscopy. Through this approach, optical techniques could detect the
predicted helical modulation in the tilt direction in the modulated state.

The ripples in tubule curvature predicted by our theory may be the
modulations seen by Yager {\it et al.}~\cite{yagerripples} and
by Thomas {\it et al.}~\cite{NoteBritt}.
Electron micrographs taken in those experiments show very clear helical
variations in the tubule curvature.  However, other experiments have not
observed any ripples on tubules, within the resolution of the electron
micrographs. The preliminary data are not yet sufficient to explain why
ripples are seen in some experiments but not in others.

Our prediction of a first-order transition between the uniform and
modulated states of tubules also has some experimental support. Nounesis
{\it et al.} have measured the magnetic birefringence and specific heat
of both single-bilayer and multi-bilayer tubules, as functions of
temperature through the melting transition~\cite{nounesis}. They find
that single-bilayer tubules undergo a second-order melting transition,
with strong pretransitional effects, while multi-bilayer tubules undergo
a first-order melting transition. Furthermore, single-bilayer tubules
show an anomalous peak in the specific heat about 3~degrees {\it below\/}
the main peak associated with melting into the untilted phase. This
anomalous peak is consistent with our predicted transition between the
modulated and the uniform states of tubules. The modulated state should
occur in the 3~degree window between the anomalous peak and the main
melting peak, where the membrane elastic constants are low, and the
uniform state should occur at lower temperatures, below the anomalous
peak. Thus, our theory may explain this heat-capacity anomaly.

\begin{figure}
\epsfxsize=6.5in\epsfbox{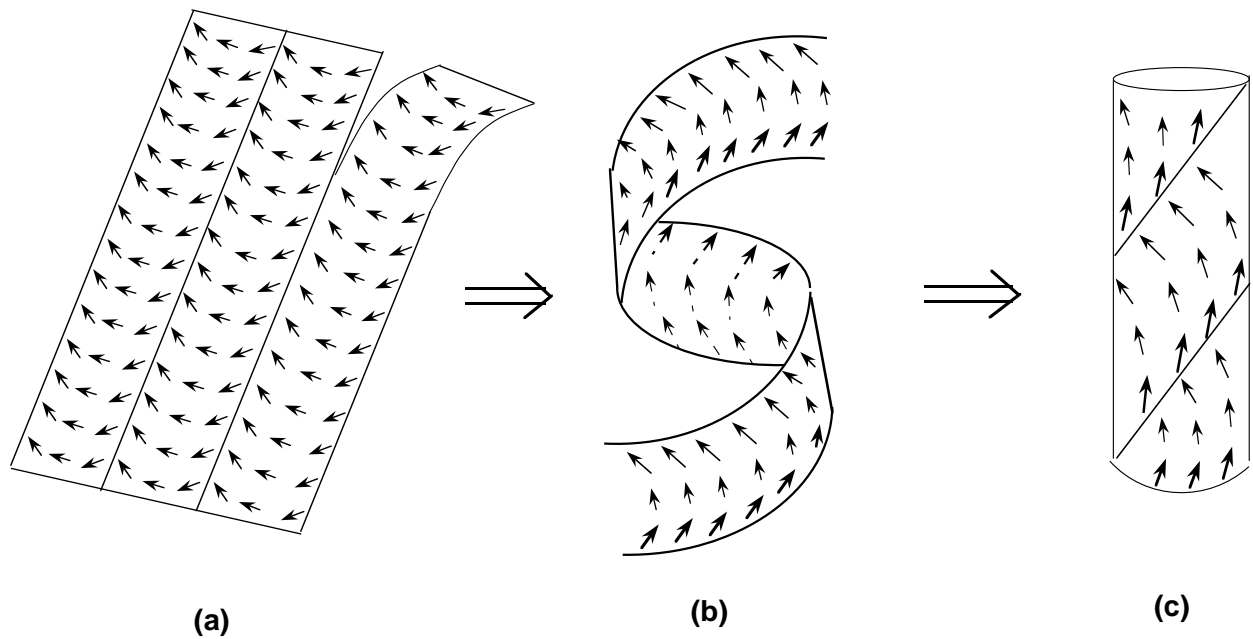}
\caption{Scenario for the kinetic evolution of flat membranes into
tubules, as discussed in the text.  (a)~When a membrane is cooled into a
tilted phase, it develops stripes in the tilt direction, and then breaks
up along the domain walls to form ribbons.  (b)~Each ribbon twists in
solution to form a helix.  (c)~A helical ribbon may remain stable, or may
grow wider to form a tubule.  Reprinted from Ref.~\protect\cite{cd} with
permission \copyright 1994 American Association for the Advancement of
Science.}
\label{kinetics}
\end{figure}

As a final point, we note that our theory for the modulated state of
tubules leads to an interesting scenario for the kinetic evolution of
flat membranes or large spherical vesicles into tubules.  This scenario,
first proposed in Ref.~\cite{cd}, is illustrated in Fig.~\ref{kinetics}.
When a flat membrane or large spherical vesicle is cooled from an
untilted into a tilted phase, it develops tilt order.  Because of the
molecular chirality, the tilt order forms a series of stripes separated
by domain walls, as shown in Fig.~\ref{kinetics}a.  Each stripe forms a
ripple in the membrane curvature, and each domain wall forms a ridge in
the membrane.  Thus, the domain walls are narrow regions where different
parts of the amphiphilic molecules come into contact with neighboring
molecules and with the solvent.  As a result, the domain walls become
weak lines in the membrane, and the membrane tends to fall apart along
those lines.  The membrane thereby forms a series of narrow ribbons.
These ribbons are free to twist in solution to form helices, as shown in
Fig.~\ref{kinetics}b.  Those helices may remain as stable helical
ribbons, or alternatively they may grow wider to form tubules, as shown
in Fig.~\ref{kinetics}c.  Note that this proposed mechanism for tubule
formation can only operate if the initial vesicle size is larger than the
favored ribbon width.  If the initial vesicle is too small, it cannot
transform into a tubule.  This prediction is consistent with the
experimental observation that large spherical vesicles (with diameter
greater than 1~$\mu $m) form tubules upon cooling, while small spherical
vesicles (diameter less than 0.05~$\mu $m) do not~\cite{rudolph,peek}.
Thus, the theoretical prediction of stripes in the tilt direction gives
some insight into the kinetics of the tubule-formation process.

In conclusion, in this paper we have shown the range of possible states
that can occur in tubules. Tubules can have a uniform state, as was
considered by earlier investigators, but they can also have a modulated
state, with a periodic helical variation in direction of molecular tilt
and in the curvature of the membrane. There is at least indirect evidence
that the modulated state occurs in actual experimental systems. A more
definitive test of this theoretical prediction requires direct
experimental probes of variations in the molecular tilt direction.

\acknowledgments
JVS and JMS acknowledge support from the Office of Naval Research, as
well as the helpful comments of G.~B.~Benedek,
D.~S.~Chung, and L.~Sloter. FCM
acknowledges partial support from the National Science Foundation under
Grant No. DMR-9257544, and from the Petroleum Research Fund, administered
by the American Chemical Society.


\end{document}